\documentclass[aps,twocolumn,preprintnumbers,showpacs,showkeys,nofootinbib%
]{revtex4}
\usepackage{epsfig}
\usepackage{amssymb,amsmath,amsfonts,amsthm,graphicx,psfrag}
\graphicspath{{./Figures/}}                 
\newcommand{\noi}{\noindent}
\newcommand{\beq}{\begin{equation}}
\newcommand{\eeq}{\end{equation}}
\newcommand{\bea}{\begin{eqnarray}}
\newcommand{\eea}{\end{eqnarray}}
\newcommand{\Fig}[1]{Fig.~\ref{#1}}
\newcommand{\Tab}[1]{Table~\ref{#1}}
\newcommand{\Sec}[1]{Section~\ref{#1}}

\newcommand{\caa}{{\cal A}}

\newcommand{\vp}{{\vec p}}

\newcommand{\tr}{\operatorname{Tr}}

\newcommand{\bc}{{\it bc~}}
\newcommand{\fc}{{\it fc~}}

\newcommand{\aleq}{\mbox{}_{\textstyle \sim}^{\textstyle < }}

\begin{document}
\preprint{ITEP-LAT/2010-10}

\title{$SU(2)$ lattice gluon propagators at finite temperatures \\
in the deep infrared region and Gribov copy effects}

\author{V.~G.~Bornyakov}
\affiliation{High Energy Physics Institute, Protvino, Russia \\
and Institute of Theoretical and Experimental Physics, 117259 Moscow, Russia}

\author{V.~K.~Mitrjushkin}
\affiliation{Joint Institute for Nuclear Research, 141980 Dubna, Russia \\
and Institute of Theoretical and Experimental Physics, 117259 Moscow, Russia}


\begin{abstract}

We study numerically the $SU(2)$ Landau gauge transverse and longitudinal
gluon propagators at non-zero temperatures $T$ both in confinement and
deconfinement phases. The special attention is paid to the Gribov copy
effects in the IR-region.  Applying powerful gauge fixing algorithm 
we find that the Gribov copy effects for the transverse propagator
$D_T(p)$ are very strong in the infrared, while the longitudinal
propagator $D_L(p)$ shows very weak (if any) Gribov copy dependence.
The value $D_T(0)$ tends to decrease with growing lattice size; however,
$D_T(0)$ is {\it non}-zero in the infinite volume limit, in disagreement
with the suggestion made in \cite{Zahed:1999tg}.  We show that in the
infrared region $D_T(p)$ is not consistent with the pole-type formula
{\it not only} in the deconfinement phase but also for $T < T_c$.
We introduce new definition of the magnetic infrared mass scale ('magnetic
screening mass') $m_M$.  The electric mass $m_E$ has been determined from
the momentum space longitudinal gluon  propagator.  We study  also the
(finite) volume and temperature dependence of the propagators as well
as discretization errors.

\end{abstract}



\maketitle

\section{Introduction}
\label{sec:introduction}

One of the most interesting features of the quantum chromodynamics at finite 
temperature is the transition from the confinement to the deconfinement
phase. This transition separates a low-temperature phase, which is
expected to be highly non-perturbative and characterized by quark and
gluon confinement, from a high-temperature - quark-gluon plasma (QGP)
- phase, where color charges should be deconfined.
The conjecture of the existence of the QGP has been supported by
recent observations of the collective effects in ultrarelativistic
heavy-ion collisions at SPS and RHIC (see, e.g., \cite{Gyulassy:2004zy}
and references therein).

\vspace{1mm}

The non-perturbative - first principle - calculation of gauge variant 
gluon (as well as ghost) propagators is of interest for
various reasons.  These propagators are expected to show different
behavior in each phase and, therefore, to serve as a useful
'order parameters', detecting the phase transition point $T_c$.
One expects that their study can shed the light on the mechanism of
the confinement-deconfinement transition.  Another reason is that
for the reliable phenomenological analysis of high-energy heavy-ion
collision data, it is important to obtain information on the momentum
dependence of the longitudinal (electric) gluon propagator $D_L(p)$
and transverse (magnetic) gluon propagator $D_T(p)$, especially in the
(deep) infrared region.  One example is the study of the radiative
energy loss in dense nuclear matter (jet quenching) which results
from the energy loss of high energy partons moving through the
plasma (see, e.g., \cite{Baier:1996kr,Baier:1996sk,Gyulassy:2003mc,
Kovner:2003zj,Wang:2000uj}). Also, the non-perturbatively calculated
lattice propagators are to be used to check the correctness of various
analytical methods in QCD, e.g., Dyson--Schwinger equations (DSE) method.
For study of the gluon propagator using DSE approach at finite temperature
see e.g. \cite{Maas:2005hs,Maas:2005ym,Cucchieri:2007ta}.  

\vspace{1mm}

The lattice study of the finite temperature $SU(2)$ gluon propagators in
Landau gauge has been performed in a number of papers (see, e.g., papers
\cite{Heller:1995qc, Heller:1997nqa,Cucchieri:2001tw, Cucchieri:2007md,
Cucchieri:2007ta,Fischer:2010fx}). 
In Ref.~\cite{Cucchieri:2001tw}
the electric and magnetic propagators were studied in both coordinate
and momentum spaces and in 4 and 3 dimensions. The conclusion was
made that the magnetic propagator had a complicated infrared behavior
which was not compatible with simple pole mass behavior. It was also
found that this propagator  had strong volume and gauge dependence.
In Refs.~\cite{Cucchieri:2007ta,Fischer:2010fx} results for both gluon and
ghost propagators in momentum space were presented, but some important
questions, e.g. Gribov copies effects, infrared behavior and scaling
were not addressed.

In paper \cite{Zahed:1999tg} it has been suggested that the proximity
of the Gribov horizon at finite temperature forces the transverse gluon
propagator $D_T(\vp,p_4=0)$ to vanish at zero three-momentum. If this
is the case, then the finite-temperature analog of Gribov formula
$~|\vp|^2/(|\vp|^4+M_M^4)~\equiv~1/(|\vp|^2+m^2_{e\! f\! f}(\vp))~$
suggests that the {\it effective} magnetic screening mass
$~m_{e\! f\! f}(\vp)~$ becomes infinite in the infrared (interpreted as
a magnetic gluons 'confinement').

The Gribov copy effects still remain one of the most serious problem in
the lattice calculations, at least, in the deep IR-region.  In our study
we employ  the gauge condition which requires the Landau gauge fixing
functional $F$ (see the definition in Section \ref{sec:definitions})
to take extrema as close as possible to the {\it global} extremum.
This choice for the gauge condition is supported by the following
facts: \noi \hspace{1mm} {\bf a)} a consistent non-perturbative gauge
fixing procedure proposed by Parrinello-Jona-Lasinio and Zwanziger
(PJLZ-approach) \cite{Parrinello:1990pm, Zwanziger:1990tn} presumes
that the choice of a unique representative of the gauge orbit should be
through the {\it global} extremum of the chosen gauge fixing functional;
\noi \hspace{1mm} {\bf b)} in the case of pure gauge $U(1)$ theory in the
weak coupling (Coulomb) phase some of the gauge copies produce a photon
propagator with a decay behavior inconsistent with the expected zero mass
behavior \cite{Nakamura:1991ww,Bornyakov:1993yy,Mitrjushkin:1996fw}.
The choice of the global extremum permits to obtain the physical -
massless - photon propagator.

In a series of papers \cite{Bakeev:2003rr,Bogolubsky:2005wf,
Bogolubsky:2007bw, Bornyakov:2008yx, Bornyakov:2009ug} we investigated the
Gribov copy effects in the Landau gauge gluon (and/or ghost) propagators
in zero temperature SU(2) gluodynamics. It has been demonstrated
unambiguously that these effects are {\it strong} in the infrared
\footnote{Unfortunately, authors of \cite{Fischer:2010fx}  cite our 
paper \cite{Bogolubsky:2005wf} in a completely misleading context.}.
Thus the Gribov copy
effects reduction is very important for the infrared behavior studies.
Recently it has been pointed out in Ref.~\cite{Boucaud:2010gr} that 
lattice results for the infrared gluon and ghost propagators free of 
Gribov copies effects would help to discriminate between scaling and 
decoupling solutions of the Dyson-Schwinger equations.
In this paper we undertake a careful study of the Gribov copy effects
in the Landau gauge gluon propagator at finite temperature. We employ
the gauge fixing procedure which we used recently in our study at zero
temperature \cite{Bornyakov:2009ug} with changes dictated by nonzero
temperature (see \Sec{sec:gribov}).

Also we attempt to make a careful analysis of (finite) volume
and temperature dependence of $D_T$ and $D_L$ as well as of scaling
violations. 

\Sec{sec:definitions} contains main definitions as well as some details
of simulations and gauge fixing procedure we use. \Sec{sec:gribov} is
dedicated to the study of the Gribov copy effects.  Volume and temperature
dependence of the propagators as well as discretization errors are
discussed in \Sec{sec:prop_mom}. \Sec{sec:screening} is dedicated
to the discussion of the screening masses and \Sec{sec:conclusions}
is reserved for conclusions and discussion.

\section{Gluon propagators: the definitions}
\label{sec:definitions}

We study the SU(2) lattice gauge theory with the standard Wilson action
\beq
S  = \beta \sum_x\sum_{\mu >\nu}
\left[ 1 -\frac{1}{2}\tr \Bigl(U_{x\mu}U_{x+\mu;\nu}
U_{x+\nu;\mu}^{\dagger}U_{x\nu}^{\dagger} \Bigr)\right], \nonumber 
\label{eq:action}
\eeq

\noi where $\beta = 4/g_0^2$ and $g_0$ is a bare coupling constant. The 
link variables $U_{x\mu} \in SU(2)$ transform  under gauge
transformations $g_x$ as follows:

\beq
U_{x\mu} \stackrel{g}{\mapsto} U_{x\mu}^{g}
= g_x^{\dagger} U_{x\mu} g_{x+\mu} \; ;
\qquad g_x \in SU(2) \; .
\label{eq:gaugetrafo}
\eeq

\noi Our calculations were performed on the asymmetric lattices with
lattice volume $V=L_4\cdot L_s^3$, where $L_4$ is the number of sites in
the $4th$ direction. The temperature $T$ is given by

\beq
T = \frac{1}{aL_4}~,
\eeq

\noi where $a$ is the lattice spacing.
We employ the standard definition of the lattice gauge vector
potential $\caa_{x+\hat{\mu}/2,\mu}$ \cite{Mandula:1987rh}:

\beq
\caa_{x+\hat{\mu}/2,\mu} = \frac{1}{2i}~\Bigl( U_{x\mu}-U_{x\mu}^{\dagger}\Bigr)
\equiv A_{x+\hat{\mu}/2,\mu}^a \frac{\sigma_a}{2} \,.
\label{eq:a_field}
\eeq

\noi The Landau gauge fixing condition is

\beq
(\partial \caa)_{x} = \sum_{\mu=1}^4 \left( \caa_{x+\hat{\mu}/2;\mu}
- \caa_{x-\hat{\mu}/2;\mu} \right)  = 0 \; ,
\label{eq:diff_gaugecondition}
\eeq

\noi which is equivalent to finding an extremum of the gauge functional

\beq
F_U(g) = ~\frac{1}{4V}\sum_{x\mu}~\frac{1}{2}~\tr~U^{g}_{x\mu} \;,
\label{eq:gaugefunctional}
\eeq

\noi with respect to gauge transformations $g_x~$.  After replacing
$U \Rightarrow U^{g}$ at the extremum the gauge condition
(\ref{eq:diff_gaugecondition}) is satisfied.

The (unrenormalized) gluon propagator $D_{\mu\nu}^{ab}(p)$ is defined 
as follows

\beq
D_{\mu\nu}^{ab}(p) = \frac{a^2}{g_0^2}
    \langle \widetilde{A}_{\mu}^a(k) \widetilde{A}_{\nu}^b(-k) \rangle
    \nonumber \\
\label{eq:gluonpropagator}
\eeq

\noi where $\widetilde{A}(k)$ represents the Fourier transform
of the gauge potentials defined in Eq.(\ref{eq:a_field}) after having
fixed the gauge, $k_i \in (-L_s/2,L_s/2]$ and $ k_4 \in (-L_4/2,L_4/2]$.
The physical momenta $p_\mu$ are given by $ap_{i}=2 \sin{(\pi k_i/L_s)}$, 
$ap_{4}=2\sin{(\pi k_4/L_4)}$.

In what follows we consider only {\it soft} modes $p_4=0$.  The hard
modes ($p_4 \ne 0$) have an {\it effective} thermal mass $2\pi T |k_4|$ 
and behave like massive particles\footnote{Let us note that the $4th$
euclidian component $p_4\ne 0$ has no physical meaning.}.

As is well known, on the asymmetric lattice there are two tensor
structures for the gluon propagator \cite{Kapusta}~:

\beq
D_{\mu\nu}^{ab}(p)=\delta_{ab} \left( P^T_{\mu\nu}(p)D_{T}(p) + 
P^L_{\mu\nu}(p)D_{L}(p)\right)\,,
\eeq

\noi where (symmetric) orthogonal projectors $P^{T;L}_{\mu\nu}(p)$ 
are defined at $p=(\vec{p}\ne 0;~p_4=0)$ as follows

\bea
P^T_{ij}(p)&=&\left(\delta_{ij} - \frac{p_i p_j}{\vec{p}^2} \right),\, 
~~~P^T_{\mu 4}(p)=0~;
\\
P^L_{44}(p) &=& 1~;~~P^L_{\mu i}(p) = 0 \,.
\eea

\noi Therefore, two scalar propagators - longitudinal $D_{L}(p)$ and
transverse $D_T(p)$ -  are given by

\bea
D_T(p)&=&\frac{1}{6} \sum_{a=1}^{3}\sum_{i=1}^{3}D_{ii}^{aa}(p)~;
\nonumber \\
D_L(p)&=& \frac{1}{3}\sum_{a=1}^{3} D_{44}^{aa}(p) \,, \nonumber \\ 
\label{gluonpropagator}
\eea

For $\vec{p} = 0$ 
propagators $D_{T}(0)$ and $D_L(0)$ are defined as follows

\bea
D_T(0) &=& \frac{1}{9} \sum_{a=1}^{3} \sum_{i=1}^{3} D^{aa}_{ii}(0)\,, 
\nonumber \\
D_L(0) &=& \frac{1}{3}\sum_{a=1}^{3} D^{aa}_{00}(0)  .
\eea

The transverse propagator $D_T(p)$  is associated  to magnetic
sector, and the longitudinal one $D_L(p)$ - to electric sector.

\vspace{2mm}

We generated ensembles of up to two thousand independent Monte Carlo
lattice field configurations. Consecutive configurations (considered as
independent) were separated by $100$ (for $L_s < 32$) or  $200$ (for $L_s
\ge 32$) sweeps, each sweep being of one local heatbath update followed
by $L_s/2$ microcanonical updates. In \Tab{tab:statistics} we provide
the full information about the field ensembles used throughout this paper.

\vspace{2mm}

In order to keep finite-volume effects under control, we considered  a 
few different lattice volumes for each temperature.  The choice of the
$6\times 48^3$ lattice at $\beta=2.635$ is important for the check of
the scaling behavior (see \Sec{sec:prop_mom}).


\begin{table}[ht]
\begin{center}
\begin{tabular}{|c|c|c|c|c|c|c|c|} \hline
 $\beta$ & $a^{-1}$[Gev] & $a$[fm] & $L_4$ & $~L_s~$ & $T/T_c$ & 
$N_{meas}$ & $N_{copy}$ \\ \hline\hline
  2.260  & 1.073 & 0.184 & 4  & 40  & 0.9     & 800     &  40     \\ 
  2.260  & 1.073 & 0.184 & 4  & 48  & 0.9     & 800     &  40     \\ \hline
  2.300  & 1.192 & 0.165 & 4  & 26  & 1.0     & 1200    &  24  \\ 
  2.300  & 1.192 & 0.165 & 4  & 40  & 1.0     & 300    &  40   \\
  2.300  & 1.192 & 0.165 & 4  & 48  & 1.0     & 400    &  40   \\ \hline

  2.350  & 1.416 & 0.139 & 4  & 16  & 1.1     & 2000    &  24     \\
  2.350  & 1.416 & 0.139 & 4  & 20  & 1.1     & 2000    &  24     \\
  2.350  & 1.416 & 0.139 & 4  & 26  & 1.1     & 1200    &  24     \\
  2.350  & 1.416 & 0.139 & 4  & 32  & 1.1     & 800     &  40     \\
  2.350  & 1.416 & 0.139 & 4  & 40  & 1.1     & 800     &  40     \\
  2.350  & 1.416 & 0.139 & 4  & 48  & 1.1     & 800     &  40     \\ \hline

  2.512  & 2.397 & 0.082 & 4  & 20  & 2.0     & 2000    &  24     \\
  2.512  & 2.397 & 0.082 & 4  & 32  & 2.0     & 800     &  40     \\
  2.512  & 2.397 & 0.082 & 4  & 40  & 2.0     & 800     &  40     \\
  2.512  & 2.397 & 0.082 & 4  & 48  & 2.0     & 800     &  40     \\ \hline

  2.635  & 3.596 & 0.055 & 6  & 40  & 2.0     & 800     &  40     \\ 
  2.635  & 3.596 & 0.055 & 6  & 48  & 2.0     & 800     &  40     \\ \hline

\end{tabular}
\end{center}
\caption{Values of $\beta$, lattice sizes, temperatures, number of
measurements and number of gauge copies used throughout this paper.
To fix the scale we take $\sqrt{\sigma}=440$ MeV. 
} 
\label{tab:statistics}
\medskip \noindent
\end{table}


\vspace{2mm}

For gauge fixing we employ the $Z(2)$ flip operation as has been proposed
in \cite{Bogolubsky:2007bw}.  It consists in flipping all link variables
$U_{x\mu}$ attached and orthogonal to a 3d plane by multiplying them
with $-1$. 

Such global flips are equivalent to non-periodic gauge transformations
and represent an exact symmetry of the pure gauge action.
The Polyakov loops in the direction of the chosen links and averaged
over the 3d plane obviously change their sign.  At finite temperature we
apply flips only to directions $\mu=1,2,3$.  In the deconfinement phase,
where the $Z(2)$ symmetry is restored, the $Z(2)$ sector of the Polyakov
loop in the $\mu=4$ direction has to be chosen 
since on large enough volumes all lattice configurations belong to the 
same sector, i.e. there are no flips between sectors.
We choose sector with positive Polyakov loop. In the
confinement phase one may use a flip in the  $\mu=4$ direction. However,
in a test run we have found that at $\beta=2.26$ studied in this paper the
maximal gauge fixing functional (\ref{eq:gaugefunctional})
has been found in the positive Polyakov 
loop sector in more than 90 \% cases. To save computer time we stick to
this sector for all configurations at this $\beta$.  Therefore, in our
study the flip operations combine for each lattice field configuration
the $2^3$ distinct gauge orbits (or Polyakov loop sectors) of strictly
periodic gauge transformations into one larger gauge orbit.

\vspace{2mm}

Following Ref.\cite{Bornyakov:2009ug} in what follows we call the
combined algorithm employing simulated annealing (SA) algorithm
(with finalizing overrelaxation) and $Z(2)$ flips the `FSA' algorithm.
For every configuration the Landau gauge was fixed $N_{copy}=24(40)$ times
($3(5)$ gauge copies for every flip--sector) on lattices with $L_s \leq 26
(L_s \geq 32)$, each time starting from a random gauge transformation of
the mother configuration, obtaining in this way $N_{copy}$ Landau-gauge
fixed copies.  We take the copy with maximal value of the functional
(\ref{eq:gaugefunctional}) as our best estimator of the global maximum
and denote it as best (``\bc'') copy.  In order to demonstrate the Gribov
copy effect we compare with the results obtained from the randomly chosen
first (``\fc'') copy.

We present results for the unrenormalized propagators and
make comments on renormalization when propagators computed at
different values of $\beta$ are compared or comparison with the
renormalized results of other groups is necessary.  Other details
of our gauge fixing procedure are described in our recent papers
\cite{Bogolubsky:2007bw,Bornyakov:2008yx,Bornyakov:2009ug}

To suppress 'geometrical' lattice artifacts, we have applied the
``$\alpha$-cut'' \cite{Nakagawa:2009zf}, i.e.  $~\pi k_i/L_s < \alpha~$,
for every component, in order to keep close to a linear behavior
of the lattice momenta $p_i \approx (2 \pi k_i)/(aL_s), ~~k_i \in
(-L_s/2,L_s/2]$.  We have chosen $\alpha=0.5$. Obviously, this cut
influences large momenta only.
We did not employ the {\it cylinder cut} in this work.

\section{Gribov copy effects and large $L_s$ behavior of $D_{T}(0)$ }
\label{sec:gribov}

\begin{figure}[tb]
\centering
\includegraphics[width=7.1cm,angle=270]{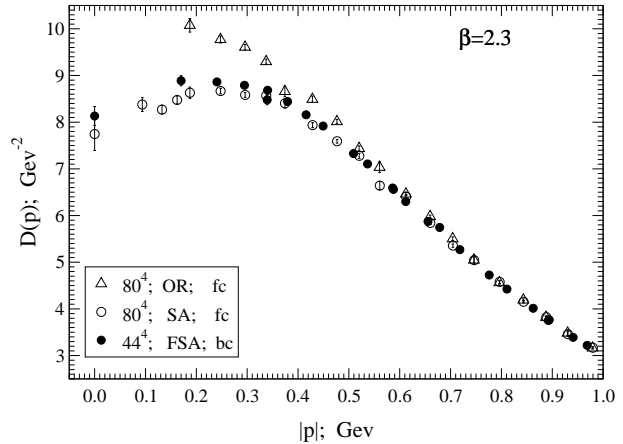}
\caption{Comparison of data obtained for
\bc FSA gauge fixing with those obtained with the standard
\fc OR - method and the \fc SA algorithm 
(this Figure from paper \cite{Bornyakov:2009ug})
}
\label{fig:OR_vs_SA}
\end{figure}

As has been already pointed above, the Gribov copy problem still remains
acute, at least, in the deep infrared region and the choice of the
efficient gauge fixing method is very important.  The importance of
this choice is demonstrated in \Fig{fig:OR_vs_SA} taken from our recent
paper \cite{Bornyakov:2009ug}. In this Figure we compare our \bc FSA
results for the bare gluon propagator $D(p)$ calculated on a $44^4$
lattice with those of the standard \fc OR method obtained for an $80^4$
lattice and also with the \fc SA results. In particular, we observe that
the OR method with one gauge copy produces completely unreliable results 
for the range of momenta $|p| \aleq 0.7$ GeV.
(The detailed discussion can be found in \cite{Bornyakov:2009ug})

\vspace{2mm}

Let us define the normalized difference of the \fc and \bc transverse
propagators $\Delta_T(p)$ :

\beq
\Delta_T(p) = \frac{D^{fc}_T(p) - D^{bc}_T(p)}
{D^{bc}_T(p)}~,
\label{delta_T}
\eeq

\noi where the numerator has been obtained by averaging over all
configurations of the difference between \fc and \bc transverse
propagators calculated for every configuration, this average being
normalized to the \bc (averaged) transverse propagator.  In a similar
way one can define also the normalized difference of the \fc and \bc
longitudinal propagators $\Delta_L(p)$.

\begin{figure}[tb]
\centering
\includegraphics[width=6.8cm,angle=270]{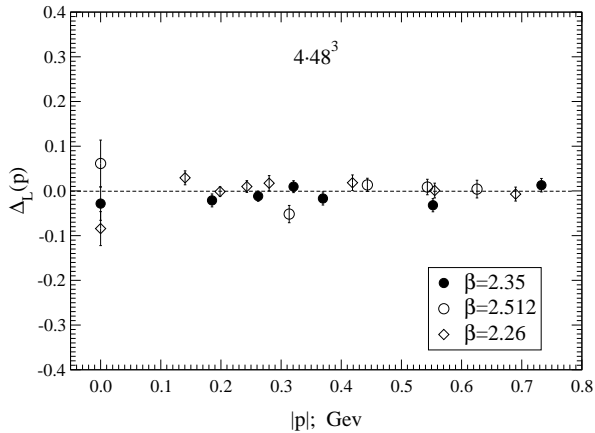}
\caption{The momentum dependence of $\Delta_L(p)$
for three temperatures. 
}
\label{fig:del_L_48x04}
\end{figure}

The longitudinal propagators $D_L(p)$ demonstrate very weak
dependence on the choice of Gribov copy. 
In Figure \ref{fig:del_L_48x04} we show the momentum dependence of
$\Delta_L(p)$ on the $4\times 48^3$ lattice at $\beta=2.35,~2.512$ and
$\beta=2.26$, i.e. at temperatures both above and below transition.  
One can see that values of $\Delta_L(p)$ are consistent
with zero for all $\beta$-values shown in the figure. This was observed
on all other lattices employed in our study.  

\vspace{1mm}

\begin{figure}[tb]
\centering
\includegraphics[width=6.cm,angle=270]{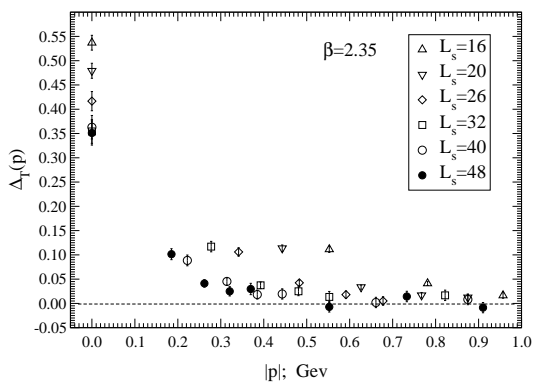}
\caption{The momentum dependence of $\Delta_T(p)$
at $\beta=2.35$ on various lattices.
}
\label{fig:del_T_2p35}
\end{figure}

In contrast, the Gribov copy dependence of the transverse propagator is
rather strong.  In Figure \ref{fig:del_T_2p35} we show the momentum
dependence of $\Delta_T(p)$ on various lattices at $\beta=2.35$
($T/T_c=1.1$). One can see that for fixed physical momentum $p$ the effect
of Gribov copies tends to decrease with increasing volume.  Such behavior
is in agreement with the absence of the Gribov copies effects (within
Gribov region) in the infinite volume limit, suggested by Zwanziger
\cite{Zwanziger:2003cf}.  On the other hand, on given lattice there
are always 3 or 4 minimal values of momentum for which these effects
are substantial. In particular, $\Delta_T(p)$ varies between $0.35$
and $0.55$ for $p=0$, between  $0.09$ and $0.12$ for $|p|=p_{min} \equiv
(2/a)\sin{(\pi/L_s)} $ and between $0.03$ and $0.04$ for momentum next
after $p_{min}$.  Similar observations were made at zero temperature as
well \cite{Bornyakov:2009ug}.

\begin{figure}[tb]
\centering
\includegraphics[width=6.cm,angle=270]{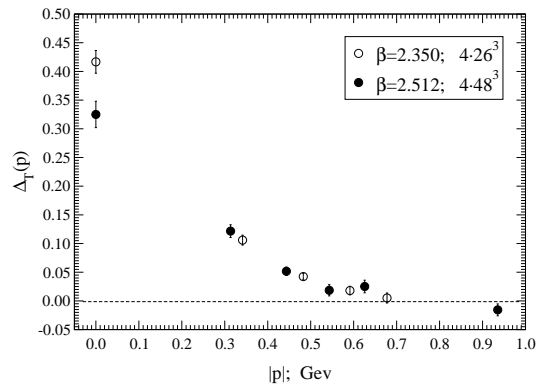}
\caption{The momentum dependence of $\Delta_T(p)$
on two lattices with approximately equal physical volume.
}
\label{fig:del_T_fixed_vol}
\end{figure}

In Figure \ref{fig:del_T_fixed_vol} we show the parameter $\Delta_T(p)$
for two temperatures, $T/T_c=1.1$ and $T/T_c=2$, on lattices with
approximately equal physical volumes. One can see that there is rather
weak (if any) dependence of the Gribov copy effects on the temperature
$T$. 

The comparison of the results for $\Delta_T(p)$ obtained on lattices $
4 \cdot 32^3$ at $\beta=2.512$ and $ 6 \times 48^3$ at $\beta=2.635$ (both
corresponding to the same temperature $T/T_c=2$ and the same physical
volume) shows that  the Gribov copy effects depend weakly on the lattice 
spacing $a$.

Let us note that results for $\Delta_T(p)$ discussed above are obtained
with the gauge fixing algorithm we have chosen, i.e., FSA. The value
of $\Delta_T(p)$ will be essentially {\it higher} if one uses the OR
algorithm to compute the \fc propagator $D^{fc}_T(p)$.

\vspace{1mm}

\begin{figure}[tb]
\centering
\includegraphics[width=6.8cm,angle=270]{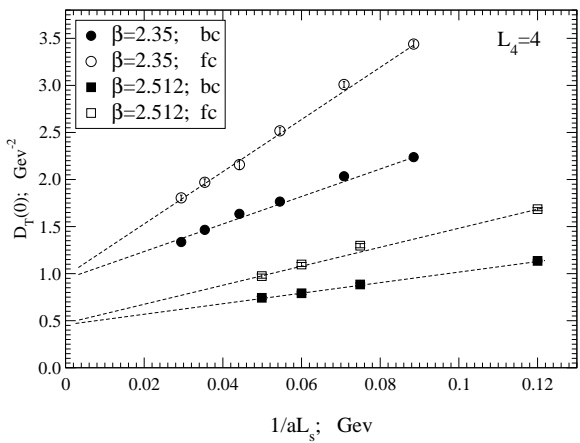}
\caption{The $1/aL_s$ dependence and Gribov copy sensitivity
of the transverse propagator $D_T(0)$ at $\beta=2.35$ (circles) 
and $\beta=2.512$ (squares). Values of $L_s$ are given in
Table \ref{tab:statistics}.
Filled symbols correspond to the \bc  ensemble, open symbols to the 
\fc ensemble.
The lines are to guide the eye.
}
\label{fig:glp0_T_2p35_2p51_bc_fc}
\end{figure}

In Figure \ref{fig:glp0_T_2p35_2p51_bc_fc} we show the $1/aL_s$ dependence
and Gribov copy sensitivity of the zero-momentum transverse propagator
$D_T(0)$ in the deconfinement phase (for $T=1.1T_c$ and for $T=2T_c$).
The difference between \bc values (filled symbols) and \fc values (open
symbols) is rather big, as has been already discussed above. However,
with increasing size $L_s$ the values of $D_T^{bc}(0)$ and $D_T^{fc}(0)$
demonstrate a tendency to decrease; moreover, 
our data, especially for $T=2T_c$, suggest that $D_T^{bc}(0)$ and
$D_T^{fc}(0)$ seem to (slowly) converge in the limit $L_s\to \infty$. This
convergence is again
in accordance with a conjecture made by Zwanziger in~\cite{Zwanziger:2003cf} 
and in accordance with the zero-temperature
case studied numerically in~\cite{Bogolubsky:2005wf,Bornyakov:2009ug}.

On the other hand, our data also suggest that $D_T(0)$ is non-zero in the 
infinite volume limit for both values of $T$, in disagreement with
the suggestion made in \cite{Zahed:1999tg}. 
There is no indication for a vanishing transverse propagator at
zero momentum for increasing volume, similarly to the situation 
for the zero-temperature case \cite{Bogolubsky:2005wf,Bornyakov:2009ug}.
This is in agreement with the refined Gribov-Zwanziger formalism 
\cite{Dudal:2007cw,Dudal:2008sp}.

Let us note that 
one still cannot exclude that there are even more efficient
gauge fixing methods, superior to the one we use, which could make this
decreasing more drastic.

\section{Volume and temperature dependence and finite 
spacing effects.}
\label{sec:prop_mom}

\subsection{Volume dependence}

\begin{figure}[tb]
\centering
\includegraphics[width=6.cm,angle=270]{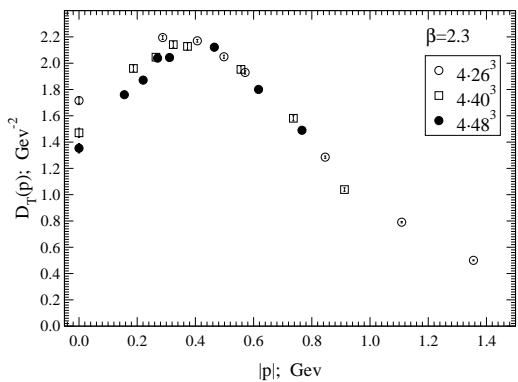}
\caption{The momentum dependence of the transverse propagators $D_T(p)$ 
on various lattices at $\beta=2.3$.
}
\label{fig:glp_T_2p30}
\end{figure}

\begin{figure}[tb]
\centering
\includegraphics[width=6.cm,angle=270]{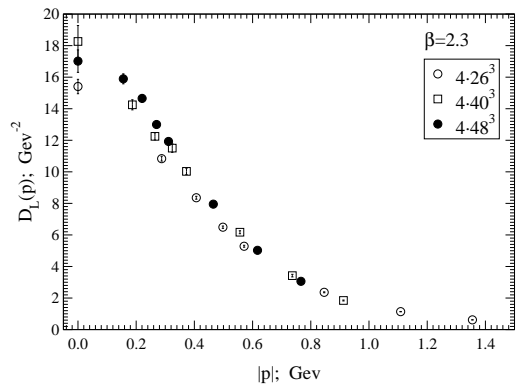}
\caption{The momentum dependence of the longitudinal propagators $D_L(p)$ 
on various lattices at $\beta=2.3$.
}
\label{fig:glp_L_2p30}
\end{figure}

As is well-known, the finite volume dependence is very strong near the
second order phase transition point $T_c$. For the transverse propagator
$D_T(p)$ this dependence can be seen from Figure~\ref{fig:glp_T_2p30}  and
for the longitudinal propagator $D_L(p)$ from Figure~\ref{fig:glp_L_2p30},
both calculated at $\beta=2.3$ (slightly above $T_c$) on various lattices.

\begin{figure}[tb]
\centering
\includegraphics[width=6.8cm,angle=270]{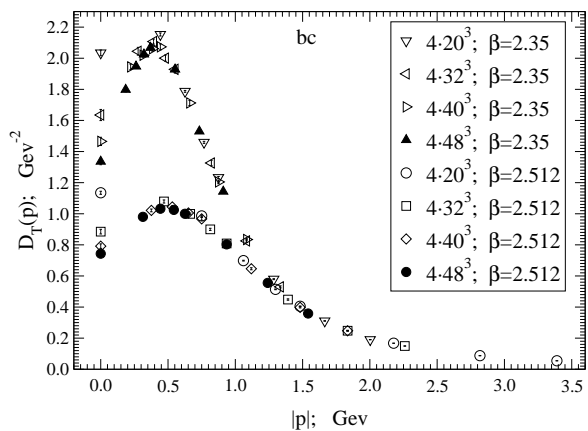}
\caption{The momentum and volume dependence of the transverse 
propagators $D_T(p)$ at $\beta=2.35$ and $\beta=2.512$.
}
\label{fig:glp_T_2p35_2p51_bc}
\end{figure}

Deeper inside in the deconfinement phase the finite volume effects
are much less pronounced, at least at non-zero values of momentum.
In Figure \ref{fig:glp_T_2p35_2p51_bc} we show the momentum dependence
of the transverse propagators $D_T(p)$ on various lattices in the
deconfinement phase at $T=1.1T_c$ and $T=2T_c$.  Apart from the strong
volume dependence at $p=0$, we see that finite volume effects are
rather weak, the sizable effects are seen only for the minimal non-zero 
momentum $p_{min}$ on a given lattice.

\begin{figure}[tb]
\centering
\includegraphics[width=6.8cm,angle=270]{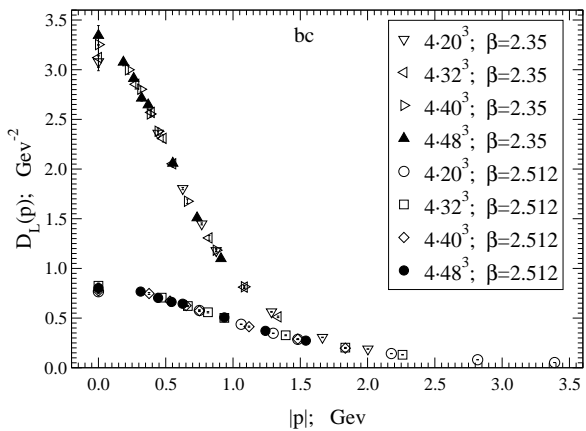}
\caption{The momentum dependence of the longitudinal propagators $D_L(p)$ 
on various lattices at $\beta=2.35$ and $\beta=2.512$.
}
\label{fig:glp_L_2p35_2p51_bc}
\end{figure}

The volume dependence of the longitudinal propagators $D_L(p)$ at 
$T=1.1T_c$ and at  $T=2T_c$ is presented in
Figure~\ref{fig:glp_L_2p35_2p51_bc}.  Evidently, the volume dependence
of $D_L(p)$ is even weaker than that of $D_T(p)$, it is rather weak
even at $p=0$. Note that at $T=1.1T_c$ $D_L(0)$ slowly increases with
increasing volume, contrary to decreasing of $D_T(0)$.

For every lattice we observed a well-pronounced maximum of the
transverse propagator $D_T(p)$ at non-zero momentum $p_0$ with
$|p_0| \sim 0.4\div 0.5$ Gev. 
Deep in the deconfinement phase the maximum has been found before  
\cite{Cucchieri:2001tw, Cucchieri:2007ta,Fischer:2010fx}.
But we observed it for the first time at $T=T_c$ and $T<T_c$
as one can see in Figure \ref{fig:glp_T_2p26_2p30_2p35_bc}.
In our recent papers \cite{Bornyakov:2008yx, Bornyakov:2009ug}
we reported about the existence of the  maximum of the scalar 
propagator $D(p)$ at non-zero momentum on the symmetric lattices
(zero-temperature case) when lattice size is big enough.
Therefore, we conclude that the transverse propagator $D_T(p)$ 
has its maximum  at $p_0\ne 0$ for {\it all} temperatures, and
the behavior of the transverse propagator in the deep infrared
is not consistent with the simple pole-type behavior both in 
confinement and deconfinement phases.

It is instructive to compare our results for $D_T(p)$ at $T=1.1T_c$ on
$L_s=48$ lattices with respective results of Ref.~\cite{Fischer:2010fx}
obtained at this temperature on the lattices with $L_4=4$ and $L_s=46$
and presented in their Figure 1. To make this comparison we made 
renormalization at $\mu=2$ GeV as it was made in \cite{Fischer:2010fx}. 
In the infrared region we found both qualitative and substantial 
quantitative disagreement between our results and results of 
Ref.~\cite{Fischer:2010fx}.  In particular, the clear maximum which
we see in our Figure~\ref{fig:glp_T_2p35_2p51_bc} at $T=1.1T_c$ can 
not be seen from Figure 1 of Ref.~\cite{Fischer:2010fx}. 
These differences between our results and results of 
Ref.~\cite{Fischer:2010fx}, which are just Gribov copies effects, are 
more essential than differences between our  \bc and \fc results 
discussed in Section~\ref{sec:gribov}. 

In contrast with $D_T(p)$, the longitudinal propagator $D_L(p)$
does not show any trace of maximum at $|p| \ne 0$ both above and
below $T_c$, as one can see in Figure~\ref{fig:glp_L_2p35_2p51_bc}
and Figure~\ref{fig:glp_L_2p26_2p30_2p35} in agreement with results of
Refs.~\cite{Cucchieri:2007ta,Fischer:2010fx}.  This gives an idea that it
can be fitted by the pole-type behavior (see Section~\ref{sec:screening}).
The pole-type behavior at high temperature is not surprising since at
high enough temperature the effective theory is the Higgs $3d$ theory
with $A_4$ playing a role of the Higgs field.

\subsection{Temperature dependence}

\begin{figure}[tb]
\centering
\includegraphics[width=6.cm,angle=270]{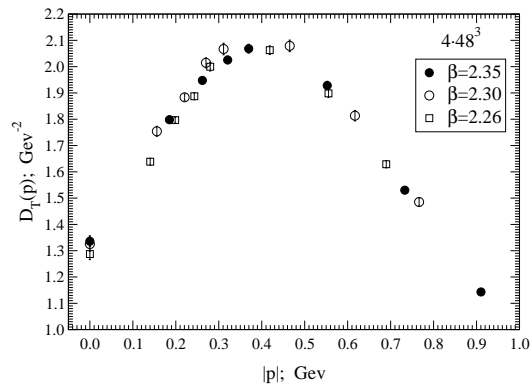}
\caption{The momentum dependence of the transverse propagators 
$D_T(p)$ near $T_c$ on the $4\times 48^3$ lattices.
}
\label{fig:glp_T_2p26_2p30_2p35_bc}
\end{figure}

The temperature dependence of the transverse propagator $D_T(p)$ 
near the critical point $T_c$ is very smooth.  Figure
\ref{fig:glp_T_2p26_2p30_2p35_bc} makes comparison of the momentum
dependence of $D_T(p)$ for three temperatures~: $T=0.9T_c$, $T=T_c$
and $T=1.1T_c$.  Indeed, there is no sign of sensitivity to the phase
transition.  It is worthwhile to note that the effect of renormalization
do not alter this conclusion since the renormalization constants computed
at $\mu=2~$GeV differ by less than 0.5\%.

Thus the transverse gluons in Landau gauge are not directly 
related to confinement \cite{Mandula:1987cp, Cucchieri:2007ta}. 

\begin{figure}[tb]
\centering
\includegraphics[width=6.cm,angle=270]{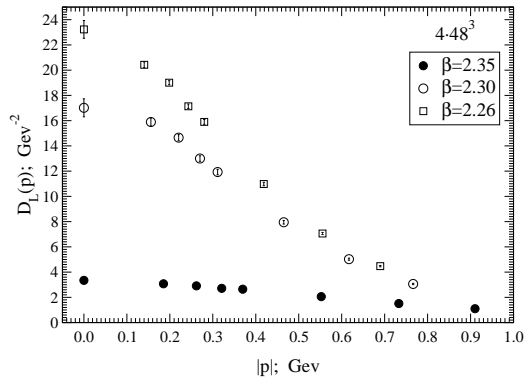}
\caption{The momentum dependence of the longitudinal gluon 
propagators $D_L(p)$ near $T_c$ on the $4\times 48^3$
lattices.
}
\label{fig:glp_L_2p26_2p30_2p35}
\end{figure}

In contrast, the longitudinal propagator  $D_L(p)$ demonstrates a
drastic jump in its values in the infrared when the critical temperature
is crossed from above (see Figure \ref{fig:glp_L_2p26_2p30_2p35}).
The respective renormalization constants computed at $\mu=2~$GeV differ
by a few percent in this case and this difference has a tendency to
decrease with increasing $\mu$. The difference in the renormalization
constants only slightly alters the temperature dependence of $D_L(p)$
after renormalization.  Therefore, $D_L(p)$  at small momenta can be 
considered as an order parameter signaling the phase transition.

Note that the study of the related quantity $\Delta_{A^2} \equiv
\langle g^2 A_E^2 - g^2 A_M^2 \rangle$ suggests that in the vicinity of
$T_c$ the temperature dependence of $D_L(p)$ can have rather nontrivial
(non-monotonous) character \cite{Chernodub:2008kf}. Further studies at $T$
very close to $T_c$ are necessary to clarify this issue.  Let us note also
that for the first time the fast change of the longitudinal propagator
near the transition point has been observed in \cite{Mandula:1987cp}
(in the $SU(3)$ case).  This fast change has been recently demonstrated
in Ref.~\cite{Fischer:2010fx} both in $SU(2)$ and $SU(3)$ theories.

\vspace{2mm}

Deep into the deconfinement phase the decreasing of the transverse
propagator $D_T(p)$ at $p\sim 0$ with increasing temperature (see
Figure~\ref{fig:glp_T_2p35_2p51_bc}) is in a qualitative agreement with
dimensional reduction since according to dimensional reduction at high
temperature $D_T(p)$ is to be proportional to $(g^2(T)~T)^{-2}$. The
quantitative agreement is not yet expected at temperatures considered here.
Similarly, the electric propagator  $D_L(p)$ for small momenta decreases
fast with increasing temperature, see Figure~\ref{fig:glp_L_2p35_2p51_bc}.

\begin{figure}[tb]
\centering
\includegraphics[width=6.8cm,angle=270]{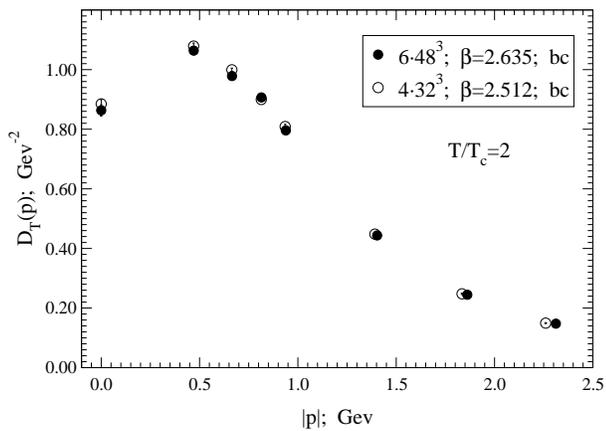}
\caption{The momentum dependence of the transverse propagator 
$D_T(p)$ on two different lattices corresponding to the same 
temperature and the same physical volume.
}
\label{fig:glp_T_2p63_2p51}
\end{figure}

\subsection{On the discretization errors}

To estimate discretization errors at $T=2T_c$  we calculated transverse and
longitudinal propagators on two different lattices corresponding
to the same physical $3d$ volume $(aL_s)^3$ but with different lattice 
spacing. These are lattices $L_4=4, L_s=32,\beta=2.512 $ and 
$L_4=6, L_s=48,\beta=2.635$. In Figure \ref{fig:glp_T_2p63_2p51} 
we show the momentum dependence of the transverse propagator $D_T(p)$ 
for these two lattices. One can see good  agreement between results 
obtained on these lattices for all included momenta.
This implies that at $T=2T_c$ the discretization effects are small even
on lattices with $L_4=4$.  We expect that this is true also at higher
temperatures.

Let us note that at this temperature we have employed larger $L_4$-values
and larger $\beta$-values (i.e., smaller spacings) as compared to that
employed in Ref. \cite{Cucchieri:2007ta}. This explains the fact that
the discretization effects we find are much smaller than in that paper.

\begin{figure}[tb] 
\centering 
\includegraphics[width=6.8cm,angle=270]{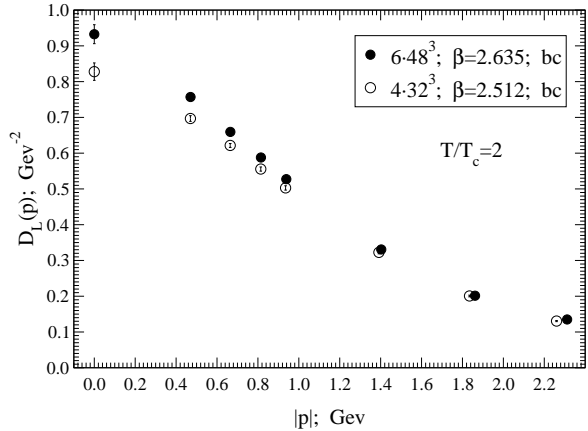} 
\caption{The momentum dependence of the longitudinal propagator 
$D_L(p)$ on two different lattices corresponding to the same 
temperature and the same physical volume.  
} 
\label{fig:glp_L_2p63_2p51} 
\end{figure} 

Contrary to the transverse propagator, the data for the longitudinal
propagator $D_L(p)$ show  substantial scaling violations in the
infrared. This can be seen from Figure~\ref{fig:glp_L_2p63_2p51} where
propagators $D_L(p)$ computed on the same lattices as used in Figure
\ref{fig:glp_T_2p63_2p51} are depicted.  The renormalization constants
computed at $\mu=2~$Gev differ only by 4\% so the scaling violations
do not disappear after renormalization.  Evidently, the discretization
errors are large at momenta smaller than 1 GeV. To reduce the finite
cut-off effects one should  increase $L_4$ or use improved lattice action
(see, e.g. \cite{Heller:1997nqa}).

The results of our study of the scaling behavior at zero temperature
\cite{Bornyakov:2009ug} suggest that for smaller values of $\beta$ (i.e.,
$\beta=2.35$, $\beta=2.3$ and $\beta=2.26$) discretization errors are
substantial for both propagators.

\section{On the screening masses}
\label{sec:screening}  

One of the interesting features of the high temperature phase is the
appearance of the infrared mass scale parameters : $m_E$ ('electric')
and $m_M$ ('magnetic'). These parameters (or 'screening masses') define
screening of electric and magnetic fields at large distances and,
therefore, control the infrared behavior of $D_L(p)$ and $D_T(p)$.
The electric screening mass $m_E$ has been computed in the leading 
order of perturbation theory long ago:  $m_E^2 = \frac{2}{3}g^2T^2$ for
$SU(2)$ gluodynamics. But at the next order the problem of the infrared
divergencies has been found.  On the other hand, the magnetic mass
$m_M$ is entirely nonperturbative in nature.  Thus a first-principles
nonperturbative calculations in lattice QCD should play an important
role in the determination of these quantities.

\begin{figure}[tb]
\centering
\includegraphics[width=6.8cm,angle=270]{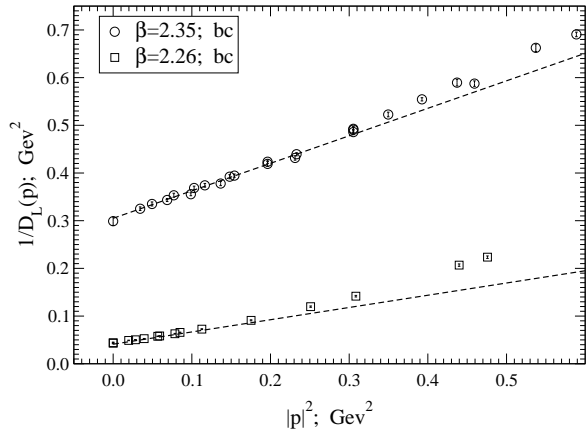}
\caption{The momentum dependence of the  inverse longitudinal propagator 
$1/D_L(p)$ at $\beta=2.35$ and $\beta=2.26$.
}
\label{fig:glpinv_L_2p35_2p26}
\end{figure}

As has been already mentioned above, the momentum dependence of the
longitudinal propagator $D_L(p)$ in the deep infrared is expected
to fit the pole-type behavior. Indeed, as an illustration,
in Figure~\ref{fig:glpinv_L_2p35_2p26} we show the momentum 
dependence of the inverse propagator $D_L^{-1}(p)$ at $T=0.9T_c$ 
and $T=1.1T_c$. Since the volume effects are small enough, 
at least, at $p\ne 0$, we use data obtained on all lattices listed 
in Table~\ref{tab:statistics} with exception for $p=0$. For this momentum 
we included data for the largest lattice only. 

One can see that at small momenta the dependence on $p^2$ is linear. 
Thus in the infrared region we have used the fitting formula

\beq
D_L^{-1}(p) = A  \cdot (p^2 + m_E^2). 
\label{eq:massfit}
\eeq

\noi The results of the fits are presented in the
Table~\ref{tab:electric_mass}. At $T=2T_c$  we can compare our results
with results of Ref.\cite{Heller:1997nqa} shown in their Figure 3. We find
that $m_E$ computed on lattices with $L_4=4$ is in good agreement with
respective result of Ref.\cite{Heller:1997nqa}.  For finer lattice spacing
($L_4=6$) our value is only slightly smaller than the value  for  $L_4=4$
(see Table~\ref{tab:electric_mass})
indicating small scaling deviations, while in  Ref.\cite{Heller:1997nqa}
the value obtained on the lattice with $L_4=8$ was substantially higher.
We believe that due to simplicity of the fitting function (\ref{eq:massfit}) 
extracting of $m_E$ from $D_L(p)$ allows more precise measurement of this 
quantity than its determination from the correlator $D_E(z)$ used in  
Ref.\cite{Heller:1997nqa}, .

In Table~\ref{tab:electric_mass} we also show the maximal momenta
$p_{max}$ included into a fit. This value was defined by condition that
the $\chi^2$/dof value for the fit was smaller than 2.
One can see that $p_{max}$ increases with $T$.


\begin{table}[ht]
\begin{center}
\begin{tabular}{|c|c|c|c|c|} \hline
$\beta$  & $T/T_c$ & $m_E$[Gev] & $m_E/T$ & $p^2_{max}$[GeV$^2$]  \\ \hline
  2.260  & 0.9     & 0.41(1) & 1.53(4) & 0.15     \\
  2.300  & 1.0     & 0.46(1) & 1.54(4) & 0.10      \\
  2.350  & 1.1     & 0.73(2) & 2.06(6) & 0.30      \\
  2.512  & 2.0     & 1.21(2) & 2.02(4) & 1.0     \\
  2.635  & 2.0     & 1.15(3) & 1.92(6) & 1.0     \\  \hline
\end{tabular}
\end{center}
\caption{Values of the screening mass $m_E$ obtained from fits 
to eq.(\ref{eq:massfit}) and maximal momenta used in the fit $p_{max}$.
} 
\label{tab:electric_mass}
\medskip \noindent
\end{table}

In contrast, the transverse propagator $D_T(p)$ has a form which is 
not compatible with the simple pole-type behavior, so for $m_M$ another, 
different from pole mass, definition is necessary. 
We applied two fitting functions to our data
for $D_T(p)$. One of them, Gaussian function with shifted argument 

\begin{equation}
 f_G(p)=C e^{-(|p|-|p_0|)^2/m_M^2}
\label{eq:fit1}
\end{equation}

\noi has been used recently in Ref.~\cite{Bornyakov:2009ug} 
to fit the $T=0$ gluon propagator in the infrared region of momenta. 
In (\ref{eq:fit1}) $m_M$ is a massive parameter, $|p_0|$ is momentum 
shift and $C$ a normalization constant. Another fitting function is 
shifted pole propagator of the form

 \begin{equation}
 f_P(p)=\frac{C}{(m_M^2+(|p|-|p_0|)^2)}
\label{eq:fit2}
\end{equation}

\noi (we keep same notations for fitting parameters).
The zero momentum was excluded from the fitting range, the maximal 
momentum was determined by requirement that respective 
$\chi^2/N_{dof}$ was smaller than 1.  
We found that both fits work well in the infrared with better performance 
(larger range) for the fit function (\ref{eq:fit2}). 


\begin{table}[ht]
\begin{center}
\begin{tabular}{|c|c|c|c|c|c|} \hline
$\beta$ & $T/T_c$ & $m_M$[Gev] & $m_M/T$ & $p_{max}$[GeV]& $|p_0|$[Gev]  \\ \hline
  2.350 & 1.1     & 0.56(1)(4) & 1.59(3)(12) & 1.3          &  0.40(1) \\
  2.512 & 2.0     & 0.78(1)(7) & 1.30(2)(11) & 1.3          &  0.51(1) \\ \hline
 \end{tabular}
\end{center}
\caption{Values of the mass parameter $m_M$ obtained from fits 
to eq.(\ref{eq:fit2}) and maximal momenta used in the fit $p_{max}$.
} 
\label{tab:magnetic_mass}
\medskip \noindent
\end{table}

In Table~\ref{tab:magnetic_mass} we show fitting parameters for fit function 
eq.~(\ref{eq:fit2}). The second error for $m_M$ is the difference from result
for fit to eq.~(\ref{eq:fit1}) in which case  $m_M$ was bigger. The difference
in values of $|p_0|$ for two fits was less than 1\%.  

We can compare our value for $m_M$ at $T/T_c=2$ with result from 
Ref.\cite{Heller:1997nqa} presented in Figure~4. In that paper 
result for $T/T_c=2$ was obtained on $32^2 \times 64 \times 8$ lattice. 
They found $m_M/T = 2.0(3)$, i.e. much higher than our value.
Apart from difference in the definition of $m_M$ this deviation might be 
explained by the Gribov copy effect, 
which is much stronger for $m_M$ than for $m_E$.
We need to make computations at higher temperatures 
to make more detailed comparison with results of 
Ref.\cite{Heller:1997nqa}.

\section{Conclusions}
\label{sec:conclusions}

In this work we studied numerically the behavior of the Landau gauge
longitudinal and transverse gluon propagators in pure gauge $SU(2)$
lattice theory in the infrared region of momentum values. The special
accent has been made on the study of the dependence of these 'observables'
on the choice of Gribov copies.

The simulations have been performed using the standard Wilson action
at temperatures from $0.9T_c$ up to $2T_c$ on lattices with $L_4=4$
and spatial linear  sizes up to $L=48$. For $T=2T_c$ simulations were
repeated on lattices with $L_4=6$. For gauge fixing gauge orbits enlarged
by $Z(2)$ flip operations were considered with up to 5 gauge copies in
every flip-sector (in total, up to $40$ gauge copies). The maximization
of the gauge functional was achieved by the simulated annealing method
always combined with consecutive overrelaxation.

\vspace{1mm}

Our findings can be summarized as follows.

\vspace{1mm}   

Similarly to the gluon propagator at $T=0$, the Gribov copy
dependence of the transverse propagators $D_T(p)$ is {\it very strong}
in the infrared, more precisely, at a few minimal (for given lattice)
momenta.  At the same time for fixed physical momentum $p$ 
the effect of Gribov copies decreases with increasing
volume in agreement with \cite{Zwanziger:2003cf}.
We found no dependence of the Gribov copies effects on the
temperature or lattice spacing.

The Gribov copy dependence of the longitudinal propagators $D_L(p)$
is very weak, at least, at non-zero momenta, and is comparable with the
statistical errors (so called 'Gribov noise').

We have to emphasize that our conclusions for Gribov copies effects are
relevant for our gauge fixing algorithm and they might change if another,
less efficient, algorithm is used.

\vspace{2mm}

With increasing size $L_s$ the  \bc-values of $D_T(0)$ and
\fc-values of $D_T(0)$ demonstrate a tendency to decrease; moreover,
$D_T^{bc}(0)$ and $D_T^{fc}(0)$ seem to (slowly) converge in the limit
$L_s\to \infty$ which is in accordance with a conjecture made by Zwanziger
in~\cite{Zwanziger:2003cf} and in accordance with the zero-temperature
case studied numerically in~\cite{Bogolubsky:2005wf,Bornyakov:2009ug}.
However, $D_T(0)$ is {\it non}-zero in the infinite volume limit, in
disagreement with the suggestion made in \cite{Zahed:1999tg}.  

\vspace{2mm}

We observed the existence of the maximum of the $D_T(p)$ at
momenta $|p| \sim 0.4\div 0.5$ Gev {\it not only} in the deconfinement
phase but also for $T \leq T_c$. Thus we confirmed that there is {\it
no} possibility to explain the IR-behavior of the transverse $D_T(p)$
gluon propagator on the basis of a simple pole-type behavior $\sim
1/(p^2 + m^2)$.  Instead we fitted this propagator to fitting functions
eq.(\ref{eq:fit1}) and eq.(\ref{eq:fit2}) with massive parameter $m_M$.
$m_M/T$ is slowly decreasing with increasing temperature. To check if
this decreasing is compatible with $gT$ behavior, as expected for the
magnetic screening mass, as well as to compare our results for this
parameter with results for magnetic screening mass, obtained by other
authors,  we need to repeat our computations at higher temperatures.

\vspace{2mm}  

For $D_L(p)$ we found good agreement with 'pole-like' behavior
at small enough momenta $p<p_{max}$ with $p_{max}$ increasing with
$T$. Our value for $m_E$ at $T=2T_c$ agrees well with result from
\cite{Heller:1997nqa} obtained also on lattices with $L_4=4$. Again,
we need results at higher temperatures to compare with other results and
with the perturbation theory predictions. We shall note that our method
of computing $m_E$ in the momentum space rather than in the coordinate
space gives rise to higher precision.

\vspace{2mm}

Away from the transition temperature the longitudinal propagators
$D_L(p)$ demonstrate very weak volume dependence.  The volume dependence
of the transverse propagators $D_T(p)$ is strong at $p=0$ and it is weak
at $p>0$.

\vspace{2mm}

We found very small scaling violations for $D_T(p)$ at $T=2T_c$
comparing results obtained on lattices with $L_4=4$ and 6. In opposite,
for $D_L(p)$ scaling violations in the infrared are substantial.

\vspace{2mm}

 We confirmed  the observation made in Ref.~ \cite{Fischer:2010fx}
that the longitudinal propagator $D_L(p)$ in the infrared increases fast
when temperature crosses transition from above. From results presented
in  Ref.~ \cite{Fischer:2010fx} for $T<T_c$ it is clear that $D_L(p)$
has a maximum near to $T_c$.  This is also in agreement with findings
for $ \Delta_{A^2} \equiv \langle g^2 A_E^2 - g^2 A_M^2 \rangle $
\cite{Chernodub:2008kf}.

\subsection*{Acknowledgments}

This investigation has been partly supported by the Heisenberg-Landau
program of collaboration between the Bogoliubov Laboratory of Theoretical 
Physics of the Joint Institute for Nuclear Research Dubna (Russia) and 
German institutes, partly by the Federal Special-Purpose Programme 'Cadres' of the Russian Ministry of Science and Education and partly 
by the grant for scientific schools NSh-6260.2010.2. VB is supported by grants 
RFBR 09-02-00338-a and RFBR 08-02-00661-a.


\end{document}